\documentclass[prl,twocolumn,epsf,psfig]{revtex4}
\usepackage{graphicx}
\DeclareGraphicsExtensions{.pdf,.png,.gif,.jpg}
\usepackage{float}
\usepackage{subfig}
\usepackage{epsfig}
\usepackage{wrapfig}
\usepackage{bbold}

\begin{document}

\title{Theoretical phase diagram of unconventional alkali-doped fullerides}

\author{Theja N. De Silva}
\affiliation{Department of Chemistry and Physics,
Augusta University, Augusta, Georgia 30912, USA;\\
Kavli Institute for Theoretical Physics, University of California, Santa Barbara, California 93106, USA.}

\begin{abstract}
By constructing an effective model based on recently calculated \emph{ab initio} bare interaction parameters, we study the phase diagram of alkali doped fullerides as a function of temperature and internal pressure. We use a slave-rotor mean-field approach at the weak and intermediate coupling limits and a variational mean-field approach at the strong coupling limit, and find a good agreement with experimental phase diagram. We explain the unified description of the phase diagram including the proximity of \emph{s}-wave superconducting state and the Mott-insulating state, and the existence of Jahn-Teller distorted metallic state using orbital selective physics. We argue that the double electronic occupation of two degenerate orbitals triggers both \emph{s}-wave superconductivity and Jahn-Teller distortion. While the orbital ordering of two electrons causes the distortion, the remaining single electron in the third orbital causes the metal-insulator transition. \end{abstract}

\maketitle

\section{I. Introduction}

Alkali-doped fullerides with the composition of A$_3$C$_{60}$, doped with A = K, Rb, Cs atoms shows the highest critical temperature about 40 K among the molecular superconductors~\cite{PI1, PI2, PI3, PI4, PI5, PI6, PI7, PI8, PI9, PI10, PI11}. Since the first discovery of the superconductivity in K$_3$C$_{60}$, the A$_3$C$_{60}$ molecular compounds gain tremendous attention recently due to their unconventional phase diagram. Fulleride compounds have been synthesized into two structures, face-centered-cubic (FCC) structure and A15 phase~\cite{PI7, PI9, PI10}. While C$_{60}$ molecules are located at the FCC positions in the FCC structure, they are located at the body-centered-cubic (BCC) positions in the A15 structure. Surprisingly, the  A$_3$C$_{60}$ molecular superconductors share a common electronic phase diagram with that of unconventional high-temperature cuprate superconductors. The superconductivity in alkali-doped fullerides (ADF's) also emerges upon chemical pressure from Mott-insulator ground state. Further, Fermi-liquid metallic phase emerges from superconducting phase upon raising the temperature. However, the alkali-doped fullerides molecular compounds considered to be unconventional with respect to the cuprate superconductors due to three main reasons. First, the phase diagram of the alkali-doped fullerides violates the common belief that the phonon-driven s-wave superconductivity and Mott-insulating state are incompatible. Second, the appearance of an unconventional Jahn-Teller metallic phase from a Jahn-Teller Mott-insulating phase upon applying internal pressure. This unconventional Jahn-Teller metallic phase shows both quasi-localized and itinerant electronic behavior simultaneously. Upon decreasing the internal pressure, this lattice distorted Jahn-Teller metallic phase makes a crossover to Fermi-liquid phase. Third, the phase diagram of the BCC structured fullerides show the evidence of co-existence of superconductivity and antiferromagnetism. These unexpected observations renew the fundamental question on the pairing mechanism of high-temperature superconductors.

The recent exciting experimental findings can be summarized as follows. At low temperatures, the A$_3$C$_{60}$ molecular compounds show superconductivity with a dome-shaped critical  temperature T$_C$ versus the lattice constant. The lattice constant of both FCC and BCC lattices of C$_{60}$ molecules, thus the volume per C$_{60}$ molecule is controlled by internal pressure with different sizes of alkali-metal ions. At higher temperatures, the compound is in either Fermi-liquid phase or Mott-insulating phase, depending on the internal pressure. The internal pressure can be quantified by the lattice constant or the C$_{60}$ molecular volume. While the compounds are conventional metals for smaller values of lattice constant, they are Mott insulators for larger values of lattice constant. By further lowering the temperature from Mott-insulating phase, the ADF's show an anti-ferro magnetic phase transition at around 2.2 and 47 K for the FCC and BCC structured fullerides, respectively~\cite{NT, NT2, NT3, NT4}. In addition, the experiments show a lattice distortion over a wide range of C$_{60}$ molecular volume at lower temperatures for the FCC lattice. The infra-red spectroscopy shows that the Jahn-Teller distortion survives well into metallic phase and at antiferromagnetic transition~\cite{PI11, Px1, Px2, Px3, Px4, Px5, Px6}.

In this paper, we propose an effective theoretical model for the ADF's. We use recent bare interaction parameters calculated from \emph{ab initio} calculations and a three-orbital Hubbard model as a basis for our proposed model. In order to study the emergence of metallic, superconducting, and Mott-insulating phases due to the competition between interaction parameters at weak and intermediate interaction regimes, we use a slave-rotor mean field theory. For the spin and orbital magnetic phase transitions at stronger interaction regimes, we use an effective spin-orbital model to construct the two-order parameter Landau energy functional. We find that the driving force behind the unconventional exotic behavior of Alkali-doped fullerides is the orbital selective physics. The renormalization of on-site interactions and Hund's coupling due to the electron-phonon interactions induces orbital rich collective phenomena. This renormalization gives orbital dependent behavior for electron dynamics. Our study shows that the orbital ordering of pairs of electron in two orbitals drive the Jahn-Teller distortion and the electron correlation in the remaining singly occupied degenerate orbital drives the Mott-insulator to metal transition. Further, we find that the emergence of s-wave superconductivity due to the local pairing of electrons in the same orbital critically depends on the Jahn-Teller phonons. Based on these theoretical observations, we conclude that the orbital selective electronic behavior due to the renormalization of on-site interaction parameters by electron-phonon coupling causes Mott-insulator, Fermi-liquid, Jahn-Teller metal, and s-wave superconducting phases proximity to be each other in Alkali-doped fullerides phase diagram.

\begin{figure}[h!]
\includegraphics[width=\columnwidth]{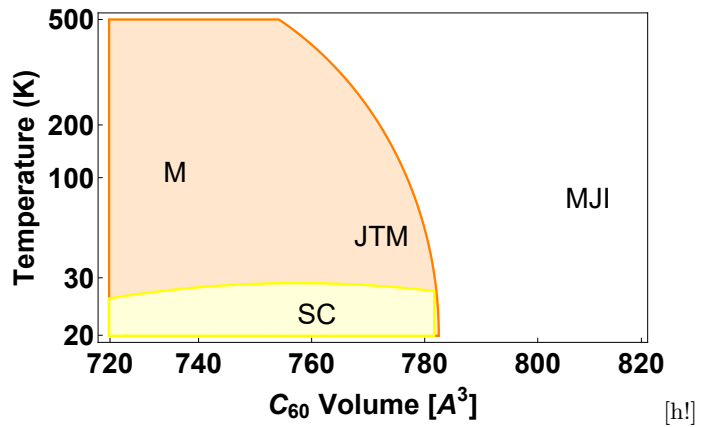}[h!]
\caption{(color online) Schematic experimental phase diagram of the FCC structured alkaline doped fullerides. The phase diagram is constructed in temperature- C$_{60}$ molecular volume space. We use the same abbreviations used by the experiments~\cite{PI11};  MJI: Mott-Jahn-Teller insulator, JTM: Jahn-Teller metal, M: conventional metal, and SC: superconductor. All solid lines are phase transitions and the transition between metal and Jahn-Teller metallic phase is a crossover.}\label{GPDxxA}
\end{figure}

\begin{figure}[h!]
\includegraphics[width=\columnwidth]{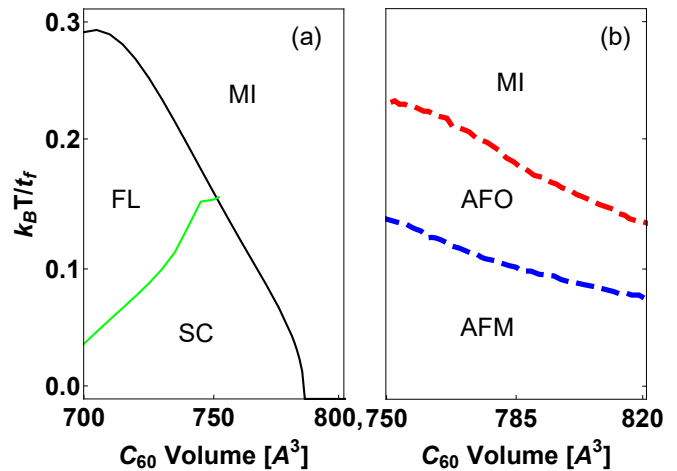}
\smallskip
\caption{(color online) Theoretical phase diagrams of the FCC structured alkaline doped fullerides for weak coupling [panel (a)] and strong coupling [panel (b)] limits. Both panels (a) and (b) has the same temperature scale in vertical axis. The phase diagrams are constructed in temperature- C$_{60}$ molecular volume space. The temperature is scaled with the hopping amplitude $t_f$ of the FCC structured fullerides. Following abbreviations are used to classify different electronic phases; MI: Mott-Insulator, FL: Fermi Liquid metal, SC: s-wave Superconductor, AFO: Antiferromagnetic orbital order, and AFM: Antiferromagnetic spin order.}\label{GPD1}
\end{figure}

We summarize our resulting theoretical phase diagram for the FCC structured fullerides in FIG.~\ref{GPD1} in comparison with a schematic experimental phase diagram in FIG.~\ref{GPDxxA}. The schematic experimental phase diagram is reproduced from the experimental results from Ref.~\cite{PI11} and Ref.~\cite{Px4}. While FIG.~\ref{GPD1} shows the theoretical phase diagram as a function of scaled temperature and volume per C$_{60}$ molecule, FIG.~\ref{GPDxxA} shows the schematic experimental phase diagram as a function of physical temperature and volume per C$_{60}$ molecule~\cite{PI11}. The theoretical phase diagrams for the BCC structured fullerides are given in section VII. Notice that we constructed the theoretical phase diagrams for the weak and strong coupling limits separately. The realistic phase diagram at intermediate C$_{60}$ molecular volume range must result from merging of these two phase diagrams. As a result, an orbital ordered Fermi liquid phase and a mixed phase of superconducting and anti-ferromagnetic can exist in the intermediate molecular volume range. If a transition from the Mott insulating phase to the orbital ordered Fermi liquid phase exists at intermediate molecular volume, it may resemble the experimentally detected phase transition from the Mott insulating phase to the Jahn-Teller metallic phase. We dedicate rest of the paper to provide detail derivation of this theoretical phase diagrams for both FCC and BCC structured alkaline doped fullerides.

\smallskip

The paper is organized as follows. In section II, starting with a three-orbital Hubbard model with electron-phonon interaction term, we construct an effective model for the ADF's. This construction is mainly based on the recent \emph{ab} initio calculation of bare interaction parameters. In section III, we use a slave-rotor approach to convert our model Hamiltonain into a coupled roton and spinon Hamiltonian. Using a mean-field theory, we decouple the roton and spinon part of the Hamiltonian and construct the phase boundaries between Fermi liquid metallic, Mott-insulating, and superconducting phases. In sections IV and V, we introduce an effective strong-coupling spin-orbital model and construct the free energy using a variational mean-field theory. In section VI, we study the spin and orbital ordering transitions by using a Landau energy functional constructed from the free energy. Finally, we present our conclusions with a discussion in section VII.

\section{II. The Effective model}

As the C$_{60}^{3-}$ molecule has a quasi-spherical structure, its electronic molecular states are spherical harmonics~\cite{Px7, Px8}. The icosahedral symmetry splits these states into icosahedral representations~\cite{Px7, Px8}. The lowest unoccupied molecular orbitals (LUMO) has three-fold degenerate orbitals. These LUMO states are separated by about 1.5 eV from highest occupied molecular orbitals (HOMO). As the next unoccupied orbital is about 1.2 eV above the LUMO, the three electrons donated by the alkali-metal ions go to the empty LUMO states.

For these multi-orbital fulleride systems, various types of interactions between electrons at C$_{60}$ molecular sites are possible. These include density-density type interactions such as on-site and off-site Coulomb interactions, and non-density type interactions, such as the pair-hopping and spin-flip interactions. In general, all these interaction terms must be included in the microscopic model. Based on \emph{ab} initio calculations~\cite{abi}, it has been shown that the off-site Coulomb interaction strength is about 25\% that of the on-site Coulomb interactions. Further, spin-flip interaction is estimated to be much smaller than that of the Coulomb interactions. It has been shown that the off-site Coulomb interaction and the spin-flip interaction do not play an essential role in driving the superconductivity~\cite{Px19}. Therefore, we neglect both off-site and spin-flip interactions in our microscopic model. Therefore by neglecting both off-site and spin-flip interactions terms, the electron dynamics of the alkali-doped fullerides can be represented by a three-orbital Hubbard Hamiltonian with an additional electron-phonon coupling term.

\begin{eqnarray}
H = \sum_{\langle ij \rangle} \sum_m \sum_{\sigma}[t_{ij} + (\epsilon_m - \mu)\delta_{ij}] a^\dagger_{im\sigma} a_{jm\sigma} \\ \nonumber +\frac{U}{2} \sum_{i} \sum_m \sum_{\sigma} n_{im\sigma} n_{im -\sigma} \\ \nonumber
+\frac{U^\prime - J}{2} \sum_{i} \sum_{m < m^\prime} \sum_{\sigma} n_{im\sigma} n_{im^\prime \sigma} \\ \nonumber
+ \frac{U^\prime + J}{2} \sum_{i} \sum_{m < m^\prime} \sum_{\sigma} n_{im\sigma} n_{im^\prime -\sigma} \\ \nonumber
+ J \sum_{i} \sum_{m \neq m^\prime} a^\dagger_{im \uparrow} a_{im^\prime \uparrow} a^\dagger_{im \downarrow} a_{im^\prime \downarrow} + H_{e-p},
\end{eqnarray}

\noindent where the electron-phonon coupling term is

\begin{eqnarray}
H_{e-p} = \sum_{i} \sum_{m m^\prime} \sum_{\nu}\sum_M \lambda_\nu V^{\nu}_{m m^\prime} a^\dagger_{im\sigma} a_{im^\prime \sigma}[b^\dagger_{i \nu M} + b_{i \nu M}].
\end{eqnarray}

Here, $a^\dagger_{im\sigma}$ ($a_{im\sigma}$) is the electronic creation (annihilation) operator in the orbital $m= 1, 2$, or $3$ with spin $\sigma = \uparrow, \downarrow$ localized at site $i$, and $n_{i m \sigma} = a^\dagger_{im\sigma}  a_{im\sigma}$ is the particle number operator. While $t_{ij}$ is the hopping integral between sites $i$ and $j$, $w =\eta t$ is the bandwidth, where the number of nearest neighbors $\eta = 12$ and $8$ for the FCC and BCC lattices, respectively. We consider only the nearest neighbor hopping $t_{ij} = t_f, t_b$ between nearest neighbor sites on the FCC and BCC C$_{60}$ molecular lattices, respectively. The on-site intra-orbital interaction $U$, inter-orbital interaction $U^\prime$, and on-site exchange interaction (bare Hund's coupling) $J$, all are related to the molecular orbital Wannier functions and the bare on-site Coulomb repulsion, as usual. The term $H_{e-p}$ represents the interaction between excess electrons  and the Jahn-Teller phonons represented by the creation operator $b^\dagger_{i \nu M}$, where $\{\nu = 0, M = 1, 2 \}$ represents 2-$A_g$ intra-molecular Jahn-Teller phonons and $\{\nu = 1 …..5, , M = 1, …..8 \}$ represents the 8-$H_g$ intra-molecular Jahn-Teller phonons~\cite{Px9, Px10, Px12, Px13}.
The intra-molecular Jahn-Teller phonon coupling constants are given by $\lambda_\nu$. Here $V^\nu_{m m^\prime}$ are the elements of coupling matrices $V^\nu$, which is determined by icosahedral symmetry~\cite{Px14, Px14B}.

The Jahn-Teller coupling induced electron-phonon coupling is extremely important for the ADF's as it favors formation of local electron-electron pairs at C$_{60}$ molecular sites~\cite{Px15, Px16}. On the other hand, the bare Hund's coupling favors high spin state at a given site. The competition between the Jahn-Teller coupling and the Hund's rule determine the total spin and the possibility of having local pairs at a molecular site. As the effective Coulomb repulsion decreases due to the electron-phonon interactions, the local electron pairing is favorable at C$_{60}$ molecular sites in ADF's. This local pairing hypothesis is confirmed by a quantum Monte-Carlo simulations for two-band degenerate orbital Hubbard model~\cite{Px17}. For the ADF's, a dynamical mean-field theory predicts local intra-orbital s-wave pairing for larger values of lattice constants at larger bare Coulomb interactions~\cite{Px18, Px19}. This double electron occupancy on each molecule at larger $U$ values, which opposed for usual Mott-Hubbard materials, is due to the lower effective interaction due to the local electron-phonon interactions. As the Hartree-Fock and many-body perturbation theories unable to predict whether ADF compunds are metallic or not, these studies conclude that the ADF compounds are on the boarder of Mott-insulator metal transition~\cite{Px20}. As the orbital degeneracy enhances the effective hopping parameter, the conductivity depends on both orbital degeneracy and the filling factors~\cite{Px21}.

In the atomic limit and the absence of Jahn-Teller electron-phonon coupling, the electron configuration of C$_{60}^{3-}$ molecule in the ADF's has three parallel electrons in LUMO states, favored by the Hund’s rule coupling. However, when the electron-phonon coupling is present, the bare interaction parameters are modified by renormalizing them due to the electron-phonon interactions. In the anti adiabatic limit, using a standard perturbation theory, the phonon variables can be eliminated and the resulting electron-phonon interaction term can be written as~\cite{Px14, Px14B},

\begin{eqnarray}
H_{e-p} \Rightarrow \frac{U_{ph}}{2} \sum_{i} \sum_{m} n_{im\uparrow} n_{im \downarrow} \\ \nonumber
+J_{ph} \sum_{i} \sum_{m \neq m^\prime} a^\dagger_{im\uparrow} a_{im^\prime \uparrow} a^\dagger_{im\downarrow} a_{im^\prime \downarrow}.
\end{eqnarray}

\begin{table}[ht]
\caption{Bare interaction parameters for the FCC structured fullerides taken from Ref.~\cite{abi}. The units are given in meV.}
\begin{center}
\begin{tabular}{|l|l|l|l|l|l|l|}
\hline
  Volume (A$^3$) & w$_f$ & U  & U$^\prime$ & J & U$_{ph}$ & J$_{ph}$ \\
\hline
 722 & 502 & 820 & 760 & 31 & -152 & -50 \\
 \hline
 750 & 454 & 920 & 850 & 34 & -142 & -51 \\
 \hline
 762 & 427 & 940 & 870 & 35 & -114 & -51 \\
 \hline
 784 & 379 & 1020 & 940 & 33 & -124 & -51 \\
 \hline
 804 & 341 & 1070 & 1000 & 36 & -134 &  -52 \\
 \hline
\end{tabular}
\end{center}
\end{table}

\begin{table}[ht]
\caption{Bare interaction parameters for the BCC structured fullerides taken from Ref.~\cite{abi}. The units are given in meV.}
\begin{center}
\begin{tabular}{|l|l|l|l|l|l|l|}
\hline
  Volume (A$^3$) & w$_b$ & U  & U$^\prime$ & J  \\
\hline
 751 & 740 & 930 & 870 & 30  \\
 \hline
 774 & 659 & 1020 & 950 & 36  \\
 \hline
 791 & 614 & 1070 & 990 & 36  \\
 \hline
 818 & 535 & 1140 & 1060 & 37  \\
 \hline
\end{tabular}
\end{center}
\end{table}

 \begin{figure}
\includegraphics[width=\columnwidth]{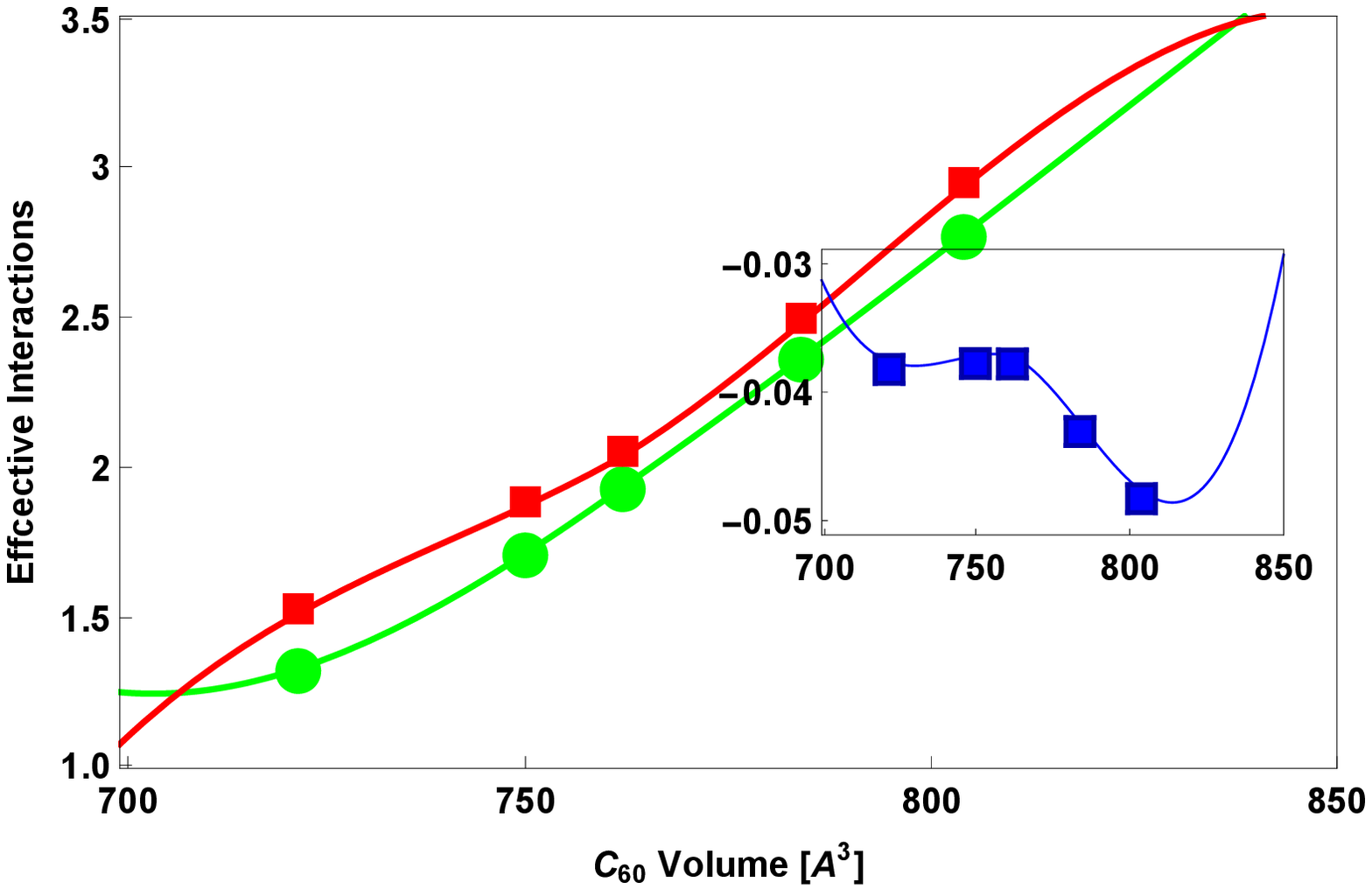}
\caption{(color online) The effective inter-orbital interaction $U^\prime/w_f$ (red, squares) and effective intra-orbital interaction $U_e/w_f$ (green, circles) as a function of the C$_{60}$ molecular volume for the FCC structured fullerides. The inset shows the effective Hund's coupling $J_{eff}/w_f$, as a function of C$_{60}$ molecular volume. The symbols are the calculation from bare interaction parameters taken from \emph{ab }initio calculations~\cite{abi}. The lines are the interpolation curves that we used in our calculations.}\label{f1}
\end{figure}

 \begin{figure}
\includegraphics[width=\columnwidth]{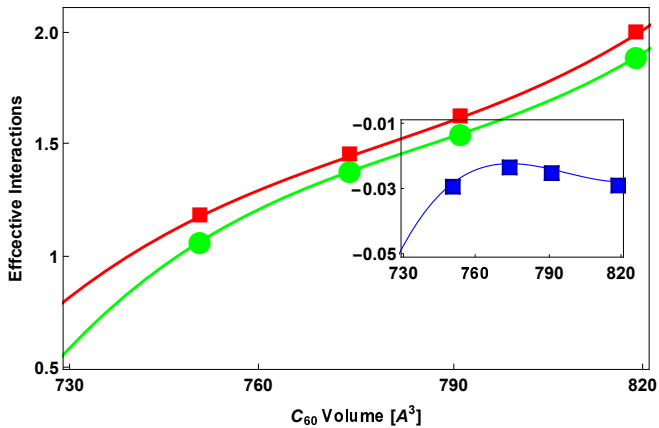}
\caption{(color online) The effective inter-orbital interaction $U^\prime/w_b$ (red, squares) and effective intra-orbital interaction $U_e/w_b$ (green, circles) as a function of the C$_{60}$ molecular volume for the BCC structured fullerides. The inset shows the effective Hund's coupling $J_{eff}/w_b$, as a function of C$_{60}$ molecular volume. The symbols are the calculation from bare interaction parameters taken from \emph{ab }initio calculations~\cite{abi}. The lines are the interpolation curves that we used in our calculations.}\label{f2}
\end{figure}

Notice that the on-site intra-orbital interaction and the bare Hund's coupling are modified by these phonon integrated electron-phonon couplings. The bare interaction parameters, $t$, $U$, $U^\prime$, $J$, $U_{ph}$, and $J_{ph}$, have already been estimated with \emph{ab} initio calculations as a function of lattice constant~\cite{abi}. These bare interaction parameters for the FCC and BCC structured fullerides are tabulated in Table.~I and Table.~II, respectively. In FIG.~\ref{f1} and FIG.~\ref{f2}, we plot the inter-orbital on-site interaction $(U^\prime)$, the effective intra-orbital on-site interaction ($U_e = U + U_{ph}$), and the effective Hund’s coupling ($J_{eff} = J + J_{ph}$) as a function of volume per C$_{60}$ molecule, taken from Refs.~\cite{abi}. Notice that the electron-phonon bare interaction parameters for the BCC structured fullerides are not available. As evidence from the Table.~I, these have relatively weak sensitivity to the C$_{60}$ molecular volume. Therefore, we use same electron-phonon interaction parameters for both FCC and BCC lattices. There are three major important observations can be drawn from FIG.~\ref{f1} and FIG.~\ref{f2}. The effective intra-orbital interaction $U_e$ decreases due to the phonon coupling ($U_e < U$) and it becomes slightly smaller than the bare intra-orbital interaction, $U_e < U^\prime$. Due to the phonon-mediated negative exchange interaction, the effective Hund's coupling becomes negative, $J_{eff} < 0 $ for entire parameter region. This inverted Hund's rule scenario for the ADF's has already been proposed before~\cite{IHR1,IHR2, IHR3}. This effective Hund's coupling favors local pairing of electrons as oppose to the high spin state. This effective parameter interaction scenario suggests that the electron configuration of a C$_{60}$ molecular site is with two-electron in one orbital and the other one is in one of the remaining two orbitals. As a result, the net spin of a C$_{60}$ molecular site is one-half. The size of the local spin-1/2 state per C$_{60}$ molecule in the Mott insulating phase is confirmed by experiments~\cite{PI8, PI10}. Notice that the effective intra-orbital interaction $U_{eff}$ curve in FIG.~\ref{f1} crosses the inter-orbital interaction $U^\prime$ at a low C$_{60}$ molecular volume for the FCC structured fullerides even without changing the sign of the effective Hund’s coupling $J_{eff}$. This seems a violation of the condition $U_{e} = U^\prime + 2J_{eff}$ below this molecular volume. This is an artifact due to our independent interpolation of the interaction parameters and the neglecting of the other non-density type interactions, such as the spin-flip term. Since the intra-orbital interaction $U^\prime$ and the effective inter-orbital interaction $U_e$ are approximately equal to each other in the entire volume per C$_{60}$ molecule range, by  approximating $U_{eff} \equiv U^\prime \simeq U_e$, the effective Hamiltonian for the ADF systems can be casted as;

\begin{eqnarray}
H = \sum_{\langle ij \rangle} \sum_m \sum_{\sigma}[t_{ij} + (\epsilon_m - \mu)\delta_{ij}] a^\dagger_{im\sigma} a_{jm\sigma}  \\ \nonumber
+ \frac{U_{eff}}{2} \sum_{i} [\sum_m a^\dagger_{im\sigma} a_{im\sigma} -\frac{N_s}{2}]^2 \\ \nonumber
+ J_{eff} \sum_{i} \sum_{m \neq m^\prime} a^\dagger_{im \uparrow} a_{im^\prime \uparrow} a^\dagger_{im \downarrow} a_{im^\prime \downarrow},
\end{eqnarray}

\noindent where $N_s = 6$ is the total number of local spin-orbital states. This is our approximated effective model for the system that includes phonon contributions. We have neglected the spin-flip and exchange terms as it has been shown that these two terms have no effect on superconductivity~\cite{Px19, np2r}. The effective interaction parameters as a function of C$_{60}$ molecular volume are given in FIG.~\ref{f1} and FIG.~\ref{f2}. The model includes three competing terms. While the hopping term favors the metallic state, the effective on-site interaction and the effective Hund's coupling compete for the Mott-insulating state and the superconducting state, respectively. Notice that the effective Hund's coupling is negative due to the phonon contributions, thus the mechanism of the on-site s-wave pairing is mediated by the phonons. The tunneling of the pairs by the negative pair-hopping interaction $J_{eff}$ in our effective model known as the Suhl-Kondo mechanism is responsible for the enhancement of superconductivity due to the inter-band scattering~\cite{R2n1, R2n2, R2n3}. In ADF's, the phonons are localized on own C$_{60}$ molecular sites and they do not propagate to neighboring molecules as much as phonon propagation in usual metal~\cite{Px14}. Even though, the momentum is well defined in the crystal, the phonons are Holstein phonons and their dispersion is flat. As a result, the conservation of momentum as the electron move through the lattice system mainly comes from the bare electrons. This indicates that the electron-phonon interaction has a weak effect on the mass renormalization of electrons. The renormalization of the electron mass in ADF's dominantly come from the Coulomb interactions. We will treat this mass renormalization through our slave-rotor approach below.

\section{III. Slave-Rotor Mean-Field Theory}

As slave-particle approaches are computationally inexpensive and capable of accounting particle correlations beyond standard mean-field theories and variational techniques, they are very popular in tackling strongly correlated particle systems. In general, the slave-particle transformation enlarges the original local Fock space of the system onto a larger local Fock space that contains more states due to the introduction of auxiliary particles. These extra nonphysical states are removed in enlarge Hilbert space by imposing constraints in an average way. Based on the studies on the first slave-particle approach~\cite{sb1}, it has been argued that slave-particle approaches are equivalent to a statistically-consistent Gutzwiller approximation~\cite{sb2, sb3, sb4}. Recently, a constraint-free, invertible canonical slave-spin transformation has been proposed for strongly correlated systems~\cite{sb5, sb6}. This slave-spin transformation is more effective than other slave-particle transformations as the basis states of the Hilbert spaces of a particle on a single site has one-to-one mapping. This one-to-one mapping excludes the additional constraint equations in this slave-spin scheme~\cite{sb6}. Instead of slave-particle or slave-spin approaches, here we use the slave-rotor approach as it's mean-field theory is economical for strongly correlated multi-orbital systems and it is constraint free at half-filling~\cite{sb7, sb8, sb9, sb10, sb11, sb12, sb13, sb14}. In this approach, the particle operator is decoupled into a fermion and a bosonic rotor that carries the spin and the charge degrees of freedom, respectively. The plan is to find a simpler description of the strongly correlated nature in terms of new effective degrees of freedom.

First, the electron operator $a_{i m \sigma}$ that annihilate an electron with spin $\sigma = \uparrow, \downarrow$ in orbital $m =1, 2, 3$ at site $i$ is expressed as a product:

\begin{eqnarray}
a_{i m \sigma} = e^{-i \theta_i} f_{i m \sigma},
\end{eqnarray}

\noindent where $f_{i m \sigma}$ represents six auxiliary fermions. This auxiliary fermion annihilates a spinon with spin $\sigma $ in orbital $m$ and the local phase degree of freedom $\theta_i$ conjugates to the total charge through the "angular momentum" operator $L_i = -i\partial/\partial \theta_i$,

\begin{eqnarray}
[\theta_i, L_j] = i \delta_{ij}.
\end{eqnarray}

\noindent In this representation, while the rotor operator $e^{-i \theta_i}$ reduces the site occupation by one unit, the eigenvalues of the $L_i$ correspond to the possible number of electrons on the lattice site. Notice that the name "angular momentum" is given due to the conservation of $O(2)$ variable $\theta_i \in [0, 2 \pi]$ but nothing to do with physical angular momentum of the electrons. Using the fact that rotons and spinons commute, one can show that the number operator of the physical particles coincide with that of spinons;

\begin{eqnarray}
n_{i m \sigma} = a^\dagger_{i m \sigma} a_{i m \sigma} = f^\dagger_{i m \sigma} f_{i m \sigma} = n^f_{i m \sigma}.
\end{eqnarray}

\noindent As the eigenvalues of the angular momentum operator $l$ can have any integer values, one must impose a constraint to truncate the enlarge Hilbert space to remove unphysical states,

\begin{eqnarray}
L_i = \sum_{\sigma, m} n^f_{i m \sigma} - 3.
\end{eqnarray}

\noindent This constraint glues charge and spin degrees of freedom and can be taken into account by introducing a Lagrange multiplier in the formalism. Notice that the angular momentum operator $L_i$ measure the particle number at each site relative to the half-filling. In terms of new variables, our Hamiltonian in Eq. (4) becomes,

\begin{eqnarray}
H = -t\sum_{\langle ij \rangle, m, \sigma}f^\dagger_{i\sigma}f_{j\sigma} e^{i(\theta_i-\theta_j)}+ \sum_{i m \sigma} (\epsilon_m - \mu - h)  f^\dagger_{im \sigma}f_{i m\sigma} \nonumber \\ +\frac{U_{eff}}{2} \sum_i L_i^2 + + J_{eff} \sum_{i, m \neq m^\prime} a^\dagger_{i m \uparrow}a_{i m^\prime \uparrow}a^\dagger_{i m \downarrow}a_{i m^\prime \downarrow},
\end{eqnarray}

\noindent where we assume nearest neighbor only hopping $t_{ij} = t$ for $i$ and $j$ nearest neighbors. Notice that the constraint is treated on average so that Lagrange multiplier $h$ is site independent. While one of the effective on-site interaction term simply becomes the kinetic energy for the rotons, the other on-site effective interaction term is still quartic. The hopping term now becomes quartic in spinon and rotor operators as well. As the slave-rotor transformation is completed, we can now decouple the effective Hamiltonian using a mean-field description~\cite{sb8}. We defined three mean-field parameters,

\begin{eqnarray}
\Delta_m = \frac{J_{eff}}{N} \sum_{i  m^\prime \neq m} \langle f_{i,m^\prime \downarrow} f_{i, m^\prime \uparrow} \rangle_f
\end{eqnarray}

\begin{eqnarray}
Q_\theta = \sum_{m \sigma} \langle f^\dagger_{i m \sigma} f_{j m \sigma} \rangle_f
\end{eqnarray}

\begin{eqnarray}
Q_f = \langle e^{i(\theta_i-\theta_j)} \rangle_\theta \equiv \langle X^\dagger_i X_j \rangle_\theta
\end{eqnarray}

\noindent where $i$ and $j$ are nearest-neighbor sites and $X_i = e^{i \theta_i}$. We will impose the condition $|X_i|^2 = 1$ using a Lagrange multiplier later. The subscript $f$ or $\theta$ means that the quantum and thermal expectation values must be taken with respect to the spinon and roton sectors, respectively. Here we make the assumptions that these expectation values are real and independent of bond directions. This mean-field decoupling allows us to transform $H \rightarrow H_f + H_\theta$, where $H_\theta$ represents an interacting quantum $XY$ model and $H_f$ represents an interacting $f$-particle spinon part. As the ADF's are half-filling electronic systems, the chemical potential $\mu = 0$ and the particle-hole symmetry requires Lagrange multiplier $h =0$. Without loss of generality, we can assume on-site energy is independent of the orbital and set $\epsilon_m = 0$, thus $\Delta_m \equiv \Delta$. The mean-field decoupling scheme leads the spinon and rotor part of the Hamiltonian to be;

\begin{eqnarray}
H_f = -t Q_f \sum_{\langle ij \rangle, m \sigma}(f^\dagger_{i m \sigma}f_{j m \sigma} + h. c) \nonumber \\  + \sum_{i m} (\Delta^\dagger_m f^\dagger_{im \uparrow} f^\dagger_{i m \downarrow} + h. c)
\end{eqnarray}

\begin{eqnarray}
H_\theta = -t Q_\theta \sum_{\langle ij \rangle}(X^\dagger_{i}X_{j} + h. c) - \lambda \sum_{i} X^\dagger_{i} X_{i}\nonumber \\ -\frac{1}{2U_{eff}} \sum_i(i \partial _\tau X^\dagger_i)(-i\partial_\tau X_i),
\end{eqnarray}

\noindent where $\lambda$ is the Lagrange multiplier to impose the condition $|X_i|^2 = 1$. Notice that the Hamiltonian is now decoupled and the mean field parameter $Q_f$ renormalizes the hopping term and related to the renormalized effective mass $m^\ast = m Q_f$. The expectation value of pairing operator, $\Delta_m$ represents the pairing of spinons. Notice that the transformed decoupled Hamiltonians posses two bosonic fields, X bosons and pair of spinons. As both of these fields can undergo Bose-Einstein condensation, it is possible to have  two global U(1) broken symmetries, one with respect to the roton field and the other with respect to the spinon field. While the metallic phase corresponds to the ordering of rotors and thus spontaneously break the O(2) symmetry, the superconducting phase corresponds to the ordering of both rotons and pair of spinons simultaneously. The simultaneous disordered rotor and the pair of spinons corresponds to the Mott-insulating phase.

As the spinon part of the transformed Hamiltonian $H_f$ is quadratic in f-fermions, it can easily be diagonalized. After performing Fourier transform into the momentum space and then usual Bogoliubov transformation, the spinon Hamiltonian $H_f$ has the form,

\begin{eqnarray}
H_f = \sum_{k,m} E_k \eta^\dagger_{k m} \eta_{k m},
\end{eqnarray}

\noindent where $\eta_{k m}$ represents the Bogoliubov quasiparticle in the spinon sector and $E_k = \sqrt{\epsilon_k^2 + \Delta^2}$ is the degenerate eigenvalues.

Here, we defined  the bare electronic energy dispersion for the electron sitting at C$_{60}$ molecular sites on the FCC and BCC lattices, $\epsilon_k = -Q_f \gamma_k$, where $\gamma_k = 4 t_f [\cos(k_xa/2)\cos(k_ya/2) + \cos(k_xa/2)\cos(k_za/2) + \cos(k_ya/2)\cos(k_za/2)]$ and  $\gamma_k = 8 t_b [\cos(k_xa/2)\cos(k_ya/2)\cos(k_za/2)]$, respectively. Notice that we set lattice constant to be $a$ and we have dropped the unimportant constant term in Hamiltonian $H_f$. The Bogoliubov quasiparticle dispersion $E_k$ is independent of the orbital index at half-filling, as we set $\epsilon_m = 0$.

The quantum and thermal expectation value of the Eq. (10) with respect to the Hamiltonian $H_f$ leads to the gap equation,

\begin{eqnarray}
\frac{1}{2J_{eff}} = -\frac{1}{N} \sum_k \frac{\tanh(\beta E_k/2)}{2 E_k},
\end{eqnarray}

\noindent where $N$ is the total number of lattice sites and $\beta = 1/k_B T$ is the dimensionless inverse temperature with Boltzmann constant $k_B$ and physical temperature $T$. Summing over nearest-neighbors and then calculating the expectation value in Eq. (11) with respect to $H_f$ gives,

\begin{eqnarray}
\eta t Q_\theta = \frac{6}{N} \sum_{k} \gamma_k \biggr[ \frac{1}{2} - \frac{\epsilon_k}{2 E_k} \tanh(\beta E_k/2) \biggr]
\end{eqnarray}

\noindent where $\eta = 12, 8$ is the number of nearest neighbor C$_{60}$  molecular sites of the FCC and BCC lattices, respectively. The expression in the square bracket is the average electronic occupation number.

The final self-consistent equation $Q_f = \langle X^\dagger_iX_j \rangle_\theta$, can easily be calculated using functional integral approach to the roton part of the Hamiltonian with the constraint equation $|X_i|^2 = 1$~\cite{sb8}. Introducing the rotor Green's function $G_\theta(k, \tau) =  \langle X_k(\tau) X^\dagger_k(0) \rangle$, the constraint equation becomes,

\begin{eqnarray}
\frac{1}{N} \sum_k \frac{1}{\beta} \sum_n G_\theta (k, i\nu_n) = 1,
\end{eqnarray}

\noindent where $\nu_n = 2n\pi/\beta$ are the bosonic Matsubara frequencies. In coherent state path integral representation, the rotor Green's can be written as,

\begin{eqnarray}
G_\theta(k, \tau) =  \frac{\int \prod_{ki}^{} \frac{dX_{ki} dX^\ast_{ki}}{2 \pi i} X(\tau) X^\ast_k(0) e^{-S_\theta}}{\int \prod_{ki}^{} \frac{dX_{ki} dX^\ast_{ki}}{2 \pi i}e^{-S_\theta}},
\end{eqnarray}

\noindent where time index $i$ labeling runs from $0$ to $\infty$ corresponding to $\tau = 0$ and $\tau = \beta$, respectively. The action in the momentum space associates with the rotor part of the Hamiltonian is given by,

\begin{eqnarray}
S_\theta = \int_{0}^{\beta} d\tau \sum_k X^\ast_k(-\frac{1}{2U_{eff}} \partial^2_\tau - \lambda - Q_\theta \gamma_k) X_k.
\end{eqnarray}

\noindent Following the standard path integral formalism, the rotor Green's function for the non-zero wave vector is given by,

\begin{eqnarray}
G_\theta (k, i\nu_n) = [\nu_n^2/U_{eff} + \lambda - Q_\theta \gamma_k]^{-1}.
\end{eqnarray}

\noindent Notice that following the Ref.~\cite{sb7}, a renormalization of $U_{eff} \rightarrow U_{eff}/2$ has been performed to preserve the exact atomic limit. Then writing,

\begin{eqnarray}
\frac{1}{\beta} \sum_n G_\theta (k, i\nu_n) = \frac{U_{eff}}{\beta} \sum_n \frac{1}{i \nu_n + \sqrt{U_{eff}(\lambda - Q_\theta \gamma_k)}} \nonumber \\ \times \frac{1}{-i \nu_n + \sqrt{U_{eff}(\lambda - Q_\theta \gamma_k)}},
\end{eqnarray}

\noindent and performing a suitable contour integration, we find

\begin{eqnarray}
\frac{1}{\beta} \sum_n G_\theta (k, i\nu_n) = \frac{U_{eff}}{2 \sqrt{U_{eff}(\lambda - Q_\theta \gamma_k})} \nonumber \\ \coth [\frac{\beta}{2} \sqrt{U_{eff}(\lambda - Q_\theta \gamma_k)}].
\end{eqnarray}

\noindent Combining this with Eq. (18) and separating $k = 0$ term in metallic phase leads the constraint equation to be,

\begin{eqnarray}
1 = Z + \frac{1}{2 N} \sum_{k} \sqrt{\frac{U_{eff}}{\lambda - Q_\theta \gamma_k}} \nonumber \\ \times \coth [\frac{\beta}{2} \sqrt{U_{eff}(\lambda - Q_\theta \gamma_k)}],
\end{eqnarray}

\noindent where $ 0 \leq Z \leq 1$ is the rotor condensate amplitude which represents the quasiparticle weight. As the rotor condensation indicates the transition into the metallic phase, non-zero quasiparticle weight $Z$ represents the metallic state. In the non-interacting limit $Z \rightarrow 1$. Finally, summing over nearest-neighbors of Eq. (12) and transforming into Fourier space leads to

\begin{eqnarray}
\eta tQ_f = \eta t Z +\frac{1}{N}\sum_{k} \frac{\gamma_k}{\beta} \sum_n G_\theta (k, i\nu_n).
\end{eqnarray}

\noindent Completing the contour integration, our final self consistent equation becomes,

\begin{eqnarray}
\eta tQ_f = \eta t Z -\frac{1}{2N}\sum_{k} \gamma_k \sqrt{\frac{U_{eff}}{\lambda - Q_\theta \gamma_k}} \nonumber \\ \times \coth [\frac{\beta}{2} \sqrt{U_{eff}(\lambda - Q_\theta \gamma_k)}].
\end{eqnarray}

\noindent This $Q_f$ is the mass enhancement factor of the quasiparticle, thus it is proportional to the effective mass of the quasiparticle $m^\ast = Q_f m$, where $m$ is the bare mass of the bare electrons. As the second term in Eq. (26) is negative, mass enhancement is always greater than the quasiparticle weight, $Q_f > Z$ at the saddle point level, and remains finite even at rotor disordered phase where $Z = 0$~\cite{sb8}. The self-consistent equations (16), (17), (24), and (26) allow us to find four unknown self-consistent parameters, $\Delta$, $\lambda$, $Q_f$, and $Q_\theta$ as a function of temperature for given interaction parameters at different C$_{60}$ molecular volume. As there are two possible global U(1) symmetry breaking for the roton and spinon sectors of the Hamiltonian, the slave-rotor theory for the ADF's predicts four distinct electronic phases. These four different phases can be characterized by two order parameters, the rotor condensate amplitude $Z$ and the spinon pairing amplitude $\Delta$. While the rotor condensate amplitude $Z$ represents the quasiparticle weight, the spinon pairing amplitude $\Delta$ represents the phase coherence of spinons. In the metallic phase rotors are condensed and the macroscopic fraction of rotor occupy the lowest energy $E_l = -\eta tQ_\theta$. As a result, the quasiparticle weight $Z$ is non-zero and the Lagrange multiplier or the rotor chemical potential $\lambda = -E_l$ is a constant in the metallic phase. The metallic electronic phase emerges in this rotor approach is the usual Fermi liquid phase. The quantum and the thermal phase transition between Fermi liquid and Mott-insulating phase is then characterized by the
vanishing quasiparticle weight $Z$ at zero spinon pairing amplitude. In the Mott-insulating phase the quasiparticle weight $Z$ is zero and the rotor chemical potential $\lambda > \eta t Q_\theta$ must be determined by self consistently. As the superconducting phase requires both metallic behavior and the phase coherence, the superconducting phase is characterized by simultaneous non zero values of $Z$ and $\Delta$. In addition to the Fermi liquid, Mott-insulating, and superconducting phases, another distinct phase can exist for $Z =0$ and $\Delta \neq 0$. This additional phase may be similar to the pseudogap phase seen in cuprate systems as it shows phase coherent, but insulating behavior~\cite{sb8}. However, we find this additional phase does not exist for the ADF systems in our present study, as it becomes anti-ferromagnetic phase when we use strong coupling model in section IV.

\begin{figure}
\includegraphics[width=\columnwidth]{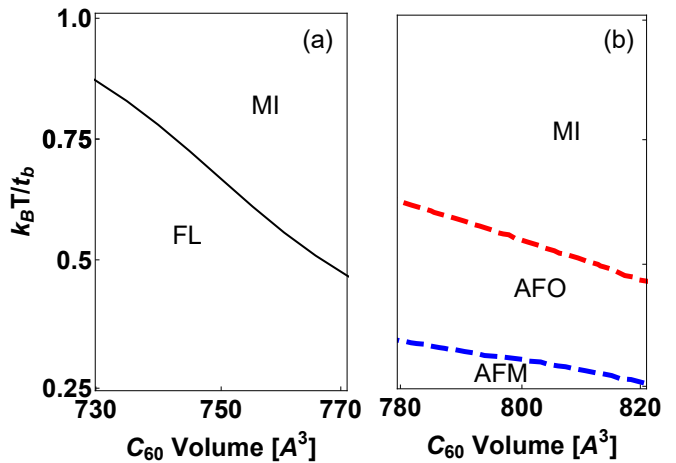}
\smallskip
\caption{(color online) Theoretical phase diagrams of the BCC structured alkaline doped fullerides for weak coupling [panel (a)] and strong coupling [panel (b)] limits. Both panels (a) and (b) has the same temperature scale in vertical axis. The phase diagrams are constructed in temperature- C$_{60}$ molecular volume space. The temperature is scaled with the hopping amplitude $t_b$ of the BCC structured fullerides. Following abbreviations are used to classify different electronic phases; MI: Mott-Insulator, FL: Fermi Liquid metal, AFO: Antiferromagnetic orbital order, and AFM: Antiferromagnetic spin order.}\label{GPDBCC1}
\end{figure}

\emph{The metal-insulator boundary line:} The Fermi liquid-Mott insulator boundary in temperature - C$_{60}$ volume parameter space is determined by setting $\Delta = 0$, $Z = 0$, and $\lambda = \eta t Q_\theta$ in self-consistent equations. We determine three unknown parameters; temperature, $Q_f$, and $Q_\theta$ from equations (17), (24), and (26). The volume dependence enters in our calculation through the interaction parameters presented in FIG.~\ref{f1} and FIG.~\ref{f2}. The solid black line shows this metal-insulator boundary line in panel (a) of FIG.~\ref{GPD1} and panel (a) of FIG.~\ref{GPDBCC1} for the FCC and BCC lattices, respectively.

\begin{figure}
\includegraphics[width=\columnwidth]{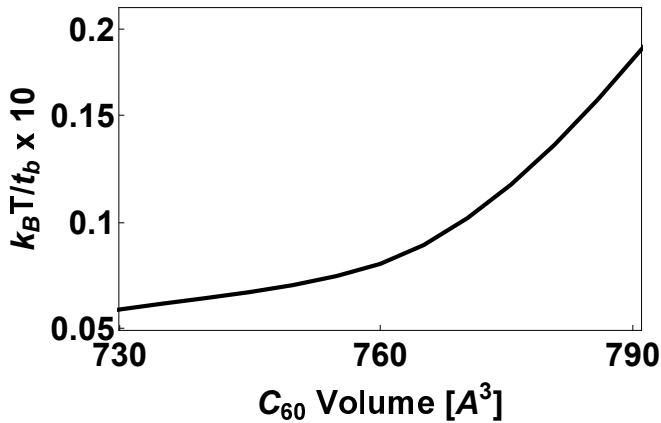}
\caption{(color online) The low temperature Fermi liquid superconductor transition for the BCC structured alkaline doped fullerides in the weak coupling limit.}\label{LTSCT}
\end{figure}

\emph{The metal-superconductor boundary line:} The Fermi liquid-superconductor boundary is determine by setting $\Delta = 0$ and $\lambda = \eta t Q_\theta$ in self-consistent equations. We then determine four unknown parameters; temperature, $Q_f$, $Q_\theta$, and $Z$ from equations (16), (17), (24), and (26). The solid green line shows this metal-superconductor boundary line in panel (a) of FIG.~\ref{GPD1} for the FCC lattice. The Fermi liquid superconductor boundary line for the BCC structured fullerides is shown in FIG.~\ref{LTSCT}.

When we solve our self-consistent equations for the boundary lines, we converted momentum summation into an integral over the first Brillouin zone by introducing a three-dimensional density of states, $D(\epsilon) = \frac{1}{N} \int \frac{d^3 k}{(2 \pi)^3} \delta(\epsilon + \gamma_k)$. The van Hove singularity of the FCC lattice at the band edge has no effect on our calculation as the integration is truncated just before the band edge. In addition, in order to construct smooth boundary lines, we use the interpolation curves for each volume in FIG.~\ref{f1} and FIG.~\ref{f2} in our calculations.

Notice that the self-consistent equations derived in this section are valid only for the weak and intermediate interaction parameters. In the following section, we will use an effective strong coupling model to construct the phase diagram at larger C$_{60}$ molecular volume.

\section{IV. Strong Interaction Limit: An effective Spin-Orbital Model}

Using a second order perturbation approach, an effective spin-orbital model has been derived for the half-filled three orbital Hubbard model relevant for the ADF compounds~\cite{som}. In this derivation, it is assumed that the exchange process for singly-occupied orbital on neighboring sites are negligible. For the case of $J_{eff} < 0$, the physics is dominated by the same orbital single occupied sites at neighboring sites. The remaining two orbitals are either empty or doubly-occupied by the other two electrons at half-filling. Therefore, a orbital pseudospin operator $\vec{\tau}_i$ has been introduced to represent the double occupation at site $i$. The orbital operator $\tau_i$ at a given site represents an orbital configuration where one orbital is doubly occupied, one orbital is singly occupied, while the remaining one is empty. Thus, $\tau^z_i = 1/2$ when the given orbital is doubly-occupied and $\tau^z_i = -1/2$ when the other one is doubly-occupied ~\cite{som}. In the derivation of effective spin-orbital model in Ref.~\cite{som}, the pair-hopping term and spin-flip terms are neglected. As a result, in the second order perturbation theory, the exchange between double-occupied and empty orbitals does not exist so that $\tau^x_i$ or $\tau^y_i$ terms do not exist in the effective spin-orbital model. Therefore, by replacing $\tau^z_i \rightarrow \vec{\tau}_i$, the effective spin-orbital model for the stronger coupling limit can be written as~\cite{som},

\begin{eqnarray}
H_{eff} = \sum_{\langle ij \rangle}[J_s \vec{S}_i \cdot \vec{S}_j + J_{\tau} \vec{\tau}_i \cdot \vec{\tau_j}\\ \nonumber
 + J_{s\tau} (\vec{\tau_i} \cdot \vec{\tau_j}) S^z_i S^z_j +\tau_{sz} S^z_i S^z_j].
\end{eqnarray}

\noindent Notice that this Hamiltonian is a simplified version of the Hamiltonian derived in Ref.~\cite{som}. Here we have neglected the coupling between projection operators $P_i$ at neighboring sites. When the coupling between projection operators at neighboring sites is negative, the expectation value of $P_i$'s are uniform. In this case, the spin-orbital structures of the system is dominated by the Hamiltonian represented by Eq. 27. Using \emph{ab} initio calculations for the bare interaction parameters~\cite{abi}, we calculate the exchange parameters derived in ref.~\cite{som} as a function of C$_{60}$ molecular volume.

For the entire molecular volume range, we find the orbital-orbital exchange parameter $J_\tau$ is larger than the spin-spin exchange parameter $J_s$ ($J_\tau > J_s$). Further, we find that the other exchange parameters ($J_{s\tau}$ and $\tau_{sz}$) are much smaller than the orbital-orbital and spin-spin exchange parameters. As $J_{s\tau}, \tau_{sz} \ll J_s, J_\tau$, the physics in the strong coupling limit determines by $J_s$ and $J_\tau$. Notice that $J_\tau > J_s$ for the entire range of C$_{60}$ molecular volume, thus one can expect orbital ordering before the spin ordering as one lowers the temperature.

Notice that the symmetry breaking of the orbital pseudo spin $\tau$ represents the orbital ordering from the orbital disordered phase to an orbital ordered phase where neighboring sites have the same single orbital occupation and one of the remaining orbitals have a double occupation. In addition to the spin sector symmetry breaking state, we are seeking for this orbital sector symmetry breaking state in the strong coupling limit. In the following two sections, we construct a two-order parameter Landau energy functional using a variational approach to study the magnetic phase diagram of the ADF molecular compounds.

\section{V. Variational Mean Filed theory at Strong Correlation Limit}

First, we use a variational mean-field approach to determine the magnetic ground state of a generic spin-orbital Hamiltonian in the form of $H_{eff}$. Once we identify the ground state magnetic structure of ADF's, we then construct a two-order parameter Landau energy functional to study the magnetic phase diagram, originating from the competition between exchange terms in the Hamiltonian $H_{eff}$ and temperature. Experimentally, a weak long ranged ordered antiferromagnetic phase has been observed below the Neel temperature $T_N \sim 2$ K for the FCC structured fullerides~\cite{PI11}. In contrast, a strong two-sublattice antiferromagnetic phase has been observed at a higher temperature $T_N \sim 47$ K for the BCC fullerides~\cite{NT2,NT3}. In order to study these magnetic transitions, we wish to break up the original C$_{60}$ molecular lattices into two sublattices $A$ and $B$, such that the molecules in each sublattices are only connected to the molecules on the other sublattice through the nearest-neighbor connections. For bipartite lattices, the molecular lattice can be divided into two sublattices such that the molecules in each sub-lattices are only connected to the molecules on the other sublattice. In this notion, the BCC lattice is a bipartite, however, the FCC lattice is traditionally non-bipartite. For the FCC lattice structure, we can still divide the lattice into two non-traditional sublattices as follows. We treat the molecules sitting at vortices of the FCC lattice belong to the $A$ sublattice and those sitting at the center of the faces belong to the $B$ sublattice. As a result, our variational mean field theory seeks for a strong two-sublattice magnetism for the BCC structured fullerides and a weak non-bipartite magnetism for the FCC structured fullerides. Once we divide the molecular lattice into two sublattices, we take our \emph{normalized variational density} matrices for the sublattices $A$ and $B$ as $\rho_{s\tau}^X = \rho_{s,X} \otimes \rho_{\tau,X}$, where

\begin{eqnarray}
\rho_{\gamma,i} = \frac{\mathbb{1}}{2} + \frac{m_\gamma}{2}(\sin \alpha_\gamma \sigma^x_\gamma \pm \sqrt{2}\cos \alpha_\gamma \cos \beta_\gamma \sigma^z_\gamma),
\end{eqnarray}

\noindent where $\sigma_\gamma$'s are components of usual Pauli matrices for spin and orbital sectors, $\gamma = s, \tau$ and $\mathbb{1}$ is the identity matrix. The upper sign is for the sublattice $i \in A $ and the lower sign is for the sublattice $i \in B$.

This choice of density matrix gives us the sublattice spin-magnetization ($m_{\gamma = s}$) and orbital-magnetization ($m_{\gamma = \tau}$) for sublattices $i \in A$ (upper sign) and $i \in B$ (lower sign), $m_{\gamma, i} = Tr(\rho_{s \tau}^i \vec{\sigma_\gamma}) \equiv \pm \xi_{\gamma \pm} \sqrt{2} m_\gamma \cos \alpha_\gamma \hat{z} + m_\gamma \sin \alpha_\gamma \hat{x}$, where $\xi_{\gamma +} = \cos \beta_\gamma$, $\xi_{\gamma -} = \sin \beta_\gamma$. Here $Tr(L)$ represents a trace of a matrix $L$. The four variational parameters $\alpha_\gamma$ and $\beta_\gamma$ are determined by minimizing the energy $E = Tr(\rho_{s,\tau} H_{eff})$. The different combination of these variational parameters provide 16 different combinations for the spin-orbital model, depending on the set of four variational values of the each sector $\gamma = s, \tau$. These 16 different combinations provide four distinct ordering patterns for each spin and orbital sectors. While $m_\gamma = 0$ represents the disordered or para-$\gamma$ ordering, non zero $\gamma$-order parameter $m_\gamma$ represents a $\gamma$-ordering phase. The three $\gamma$-ordered pases are $XY$-ferromagnetic-$\gamma$ ordering ($\alpha_\gamma = \pi/2$), z-antiferromagnetic-$\gamma$ ordering ($\alpha_\gamma = 0$ and $\beta_\gamma = \pi/4$), and z-ferromagnetic-$\gamma$ ordering ($\alpha_\gamma = 0$ and $\beta_\gamma = -\pi/4$). The values of the variational parameters depend on the ground state energy that determines by the exchange interaction parameters in the effective spin-orbital Hamiltonian $H_{eff}$.

\begin{eqnarray}
E = \frac{N \eta}{8} \biggr[J_s m_s^2 [\sin^2 \alpha_s - \cos^2 \alpha_s \sin(2\beta_s)] \\ \nonumber
+ J_\tau m_\tau^2 [\sin^2 \alpha_\tau - \cos^2 \alpha_\tau \sin(2\beta_\tau)]  \\ \nonumber
- J_{sz} m_s^2 \cos^2 \alpha_s \sin (2 \beta_s) \\ \nonumber
+ \frac{J_{s\tau}}{4} m_s^2 m_\tau^2 \cos^2 \alpha_s \cos^2 \alpha_\tau \\ \nonumber
\times \sin(2\beta_s) \sin(2 \beta_\tau) \biggr].
\end{eqnarray}

\noindent The entropy contribution to the free energy $-TS = k_BT \rho_{s, \tau} \ln \rho_{s, \tau} \equiv k_BT \sum_{a_\gamma} \lambda_{a_\gamma}\ln \lambda_{a_\gamma}$, where the four eigen values of the density matrix is given by,

\begin{eqnarray}
\lambda_{a_\gamma} = \frac{1}{2} \pm \frac{m_\gamma}{2} \sqrt{1 + \cos^2 \alpha_\gamma \cos (2 \beta_\gamma)}.
\end{eqnarray}

\noindent In general, the minimization of Helmholtz free energy $F = E-TS$ for given exchange interaction parameters allows one to determine the variational parameters, thus the spin and orbital structures on the FCC and BCC lattices.

\section{VI. Landau Energy Functional for Magnetic and orbital Orders}

For the exchange interaction parameters relevant for the FCC and BCC structured fullerides, we find that the ground state of ADF compounds are antiferromagnetic-spin (AFM-S) and antiferromagnetic-orbital (AFM-O). The ground state energy for this both spin and orbital AFM ordered state is given by,

\begin{eqnarray}
E = \frac{N \eta}{8} \biggr[-(J_s + J_{sz}) m_s^2 - J_\tau m_\tau^2 + \frac{J_{s\tau}}{4} m_s^2 m_\tau^2 \biggr].
\end{eqnarray}

\noindent The entropy contribution to the free energy $-TS =  k_BT \sum_{a_\gamma} \lambda_{a_\gamma}\ln \lambda_{a_\gamma}$ is then determines using the eigen values of the density matrix, $\lambda_{a_\gamma} = \frac{1}{2} (1 \pm m_\gamma)$. The free enegy $F = E-TS$ is now a function of two-order parameters, $m_s$ and $m_\tau$ for AFM-spin order and AFM-orbital order. In order to find the critical temperatures for the spin and orbital ordering, we construct a Landau energy functional by expanding the free energy in powers of $m_s$ and $m_\tau$ and keeping only the powers upto quartic order. The two-order parameter Landau energy functional or the free energy per site $f = F/N$ up to the quartic order is then given by,

 \begin{eqnarray}
f = \frac{1}{2} A_s m_s^2 + \frac{1}{2} A_\tau m_\tau^2 + \frac{1}{4} B_s m_s^4 \\ \nonumber
+ \frac{1}{4} B_\tau m_\tau^4 + \frac{1}{2} G m_s^2 m_\tau^2,
\end{eqnarray}

\noindent where the temperature dependent constants are given by $A_s = -\eta (J_s + J_{sz})/4 + 2k_BT$, $A_\tau = -\eta J_\tau/4 + k_BT$, $B_s = B_\tau = 2k_BT/3$, and $G = \eta J_{s\tau}/16$. The spin-orbital phase diagram can be constructed analytically by minimizing the Landau energy functional for order parameters $m_s$ and $m_\tau$. By defining three new variables, $X_s = A_s/\sqrt{B_s}$, $X_\tau = A_\tau/\sqrt{B_\tau}$, and $\Lambda = G/\sqrt{B_s B_\tau}$, we find the paramagnetic phase is stable when both $X_s > 0$ and $X_\tau > 0$. The AFM-spin and paramagnetic-orbital phases require $X_s$ to be negative. Similarly, the AFM-orbital and paramagnetic-spin phases require $X_\tau$ to be negative. For the AFM-S and para-O phase requires $\Lambda > X_\tau/X_s$ or $\Lambda < max(-1, X_s/X_\tau)$ when $X_\tau > 0$ and $\Lambda < min(X_s/X_\tau, X_\tau/X_s)$ or $\Lambda  < -1$ with $X_s^2 > X_\tau^2$ when $X_\tau < 0$. Likewise, for the para-S and AFM-O phase requires $\Lambda > X_s/X_\tau$ or $\Lambda < max(-1, X_\tau/X_s)$ when $X_s > 0$ and $\Lambda < min(X_s/X_\tau, X_\tau/X_s)$ or $\Lambda  < -1$ with $X_s^2 < X_\tau^2$ when $X_s < 0$. If non of the above conditions satisfy, we find both AFM-spin and AFM-orbital phases, simultaneously. The simultaneous spin and orbital ordered phase requires both or one of the parameters $X_s$ or $X_\tau$ to be negative. Using this criteria for the exchange interaction parameters, we construct the magnetic phase boundaries for the ADF's. These boundary lines are shown in panel (b) of FIG.~\ref{GPD1} and FIG.~\ref{GPDBCC1}. While the dashed red line shows the para magnet to antiferro-orbital transition, the dashed blue line represents the transition into antiferro-spin state.

\section{VII. Conclusions and Discussion}

The phase diagrams obtained by solving the weak/intermediate coupling effective Hamiltonian and the strong coupling effective Hamiltonian are shown in FIG.~\ref{GPD1}, FIG.~\ref{GPDBCC1}, and FIG.~\ref{LTSCT}. Our phase diagrams show all the features seen in experimental phase diagrams~\cite{PI8, PI11, NT2, NT3}.

As we have discussed above, both s-wave superconductivity and orbital ordering originate from the double occupation of electrons in two-degenerate orbital. The metal-insulator behavior originates from the singly occupied electron in the remaining orbital. As a result, the metallic behavior can survive even below the orbital ordering temperature. Even though, ADF compounds show Jahn-Teller distortion, there is no evidence for structural transition across the metal-insulator transitiion~\cite{PI7,PI8, PI10}. Based on our understanding on transition metal oxides, when there is an orbital ordering, it accompanies by the Jahn-Teller distortion~\cite{JTO1}. The two co-operative phenomena in transition metal oxides; electronic orbital ordering and structural Jahn-Teller distortion are concurrent~\cite{JTO2, JTO3}. In other words, both Jahn-Teller distortion and orbital ordering occurs simultaneously in transition metal oxides. It is therefore, extremely difficult to distinguish the cause and effect. The parent compound of the colossal magneto-resistance manganites LaMnO$_3$ and the cubic perovskites KCuF$_3$ are considered as two classic text-book examples for these co-operative simultaneous phenomena~\cite{JTOO1, JTOO2, JTOO3, JTOO4}. Here we argue that the Jahn-Teller distortion detected in experimental phase diagram is a result of the double electron orbital ordering in FCC ADF’s. However, unlike transition metal oxides, single occupied electron in one of the orbitals does not participate in the orbital ordering.  Therefore, the ADF systems can show metallic behavior in the Jahn-Teller distorted phase too. In this notion, the Jahn-Teller metallic phase in experimental phase diagram resembles our antiferro-orbital ordered metallic phase. As our strong coupling theory is not valid for the weak coupling regime, we are unable to find the crossover transition from the orbital ordered metallic phase to the usual Fermi liquid phase. In a different note, one can argue that the interpretation of Jahn-Teller metal phase is due to the fact that the time scale of lattice dynamics. When the time scale of lattice dynamics becomes slow, the molecules can look distorted within the time scale of experimental probe. In addition to antiferro-orbital order, we find antiferro-spin ordering transition at a lower temperature. As a result, the ADF systems at stronger coupling limit is in a both spin and orbital ordered phase. Experimentally, the antiferromagnetism for the FCC ADF's has been observed at extremely low temperatures~\cite{NT}. Even below the Neel temperature $T_N \sim 2$ K, the specific heat measurements suggest that the suppression of antiferromagnetism, thus the weak long-range magnetic order~\cite{SPH}. This suppression of Neel's order may be partly attributed to the frustration caused by the non-bipartite nature of the FCC lattice and the orbital ordering. As our sublattice division for the FCC lattice is not a traditional sublattice division due to the non-bipartite nature of the FCC lattice, the antiferromagnetism in our theory is a weak one, but not well defined two-sub lattice antiferromagnetism. On the other hand, the well defined two sublattice antiferromagnetism for the BCC ADF's has been observed at a higher temperatures~\cite{NT2, NT3}. This observation is consistent with our strong coupling theory for the BCC structured ADF's.

Surprisingly with our mean-field approaches, we find that when the s-wave superconducting states ends at the metal-insulator transition line for the FCC structure, the system becomes orbital ordered antiferromagnet. Notice that these two different states are discovered by using two different effective Hamiltonians with different approximate theoretical approaches. As our theoretical scheme is incapable of detecting, we cannot rule out the simultaneous existence of superconductivity and antiferromagnetism around the superconductor-AFM transition line. Indeed, the experimental results suggest that the simultaneous co-existence of antiferromagnetism and superconductivity in the BCC structured fullerides~\cite{PI8, NT2}. Further, based on our investigation, it is not surprising to see both s-wave superconductivity and Mott-insulator proximity in the phase diagrams. This is because these two phases emerge due to the electron occupation of orbital selective scenarios. Using our effective theoretical model and its solution, we were able to recover all four electronic phases discovered in experimental phase diagram~\cite{PI11}. In addition to the common Fermi liquid metallic phase and s-wave superconducting phase, the experimental Jahn-Teller metallic (JMT) and Jahn-Teller Mott-insulating (MJI) phases resemble our orbitally ordered Fermi liquid phase and orbitally ordered Mott-insulator, respectively. Moreover, we discover two more additional phases in the temperature-C$_{60}$ molecular volume space. The high-temperature Mott-insulator (MI) and the low-temperature orbitally ordered antiferromagnetic (AFM) phases discovered in our theoretical phase diagram have not been extensively investigated in experimental phase diagrams~\cite{PI8, PI11, NT2}. However, it is not surprising to expect high-temperature regular Mott-insulator for the ADF compounds. The low-temperature antiferromagnetic phase at larger C$_{60}$ molecular volumes has already been detected in experiments~\cite{NT, NT2, NT3}. Notice that we have neglected the possible charge density wave stabilization in our study as the strong on-site Coulomb repulsion dominates over the weak off-site Coulomb repulsion in ADF’s. The absence of charge density wave state is experimentally confirmed by the NMR studies~\cite{NT2}. Further, the absence of the instability in the charge sector can be justified by the existence of Mott insulating phase due to the strong on-site Coulomb repulsion where the charge degrees of freedom is frozen and the total electronic occupation per site is fixed to be at three. As the effective interorbital interaction U$^\prime$ is greater than the effective intraorbital interaction U$_e$, the ADF compounds show orbital instability.

The boundary lines between different phases in temperature-C$_{60}$ molecular volume space are also in qualitative agreement with experiments. Notice that the we have scaled temperature in our theoretical phase diagrams with the hopping amplitudes $t_f$ and $t_b$ for the FCC and BCC structured fullerides, respectively. In addition to the Coulomb interaction parameters, the hopping amplitudes also changes upon applying the internal pressure~\cite{abi}. This is the reason why we have constructed our theoretical phase diagrams as a function of dimensionless temperature $k_BT/t_f$ and $k_BT/t_b$. In order to a get a quantitative comparison, we use few specific known $t_f$ values for the FCC structured fullerides from ref.~\cite{abi} and calculate the physical temperatures on the metal-insulator line. We find our theory slightly over estimate the transition temperature. For example, the metal-insulator transition occurs at about 860 K and 563 K at the C$_{60}$ molecular volumes 750 A$^3$ and 762 A$^3$, respectively. The overestimation of the theoretical metal-superconductor transition temperature is bit higher than that of the metal-insulator transition temperature. Several factors, such as the crudeness of mean-field theories, the re-normalization of electron mass due to the phonon contribution which we assumed to be negligible, and the longer-range orbital dependent hopping may be attributed to this overestimation. Regardless the quantitative agreement of the critical temperature, we manage to recover all other features in experimental phase diagrams using our effective theoretical model. We are unable to make a quantitative comparison for the BCC structured fullerides as some bare interaction parameters, such as phonon mediated couplings are not known. Note that we use the same phonon mediated coupling parameters for both FCC and BCC structured fullerides. Therefore, our weak coupling phase diagram for the BCC structured fullerides must be considered as qualitative.

In conclusion, by using recently calculated \emph{ab} initio interaction parameters we have proposed an effective theoretical model for the alkali-doped fullerides compounds. Employing a slave-rotor and an effective spin-orbital mean field theories, we recovered all the dominant features in experimental phase diagrams. We find that the proximity of various electronic phases including the Mott-insulator, the s-wave superconductor, and the Fermi liquid and the existence of Jahn-Teller metallic phase trigger from the orbital selective electronic occupations.

\section{VIII. Acknowledgments}

The author acknowledges the support of Augusta University and the hospitality of KITP at UC-Santa Barbara. A part of this research was completed at KITP and was supported in part by the National Science Foundation under Grant No. NSF PHY11-25915.

\end{document}